\documentclass[11pt]{article}
\usepackage[latin1]{inputenc}
\usepackage{amsmath}
\usepackage{amsfonts}
\usepackage{amssymb}
\usepackage{color}
\usepackage{epsfig}
\newlength{\bredde}
\def\slash#1{\settowidth{\bredde}{$#1$}\ifmmode\,\raisebox{.15ex}{/}
\hspace*{-\bredde} #1\else$\,\raisebox{.15ex}{/}\hspace*{-\bredde} #1$\fi}
\newcommand{\noi}{\vspace{12pt}\noindent}

\newcommand{\be}{\begin{equation}}
\newcommand{\ee}{\end{equation}}
\newcommand{\bea}{\begin{eqnarray}}
\newcommand{\eea}{\end{eqnarray}}
\newcommand{\nn}{\nonumber}
\newcommand{\ow}{\overline w}
\newcommand{\oer}{\overline r}
\newcommand{\oq}{\overline q}
\newcommand{\ok}{\overline \kappa}
\newcommand{\OO}{\overline \Omega}
\newcommand{\oS}{\overline S}

\newcommand{\one}{\mbox{\bf 1}}
\newcommand{\bs}{\begin{split}}
\newcommand{\es}{\end{split}}

\newcommand{\sect}[1]{\setcounter{equation}{0}\section{#1}}

\def\Tr{{\mbox{Tr}}}
\def\Pf{{\mbox{Pf}}}

\def\im{{\Im\mbox{m}}}
\def\aa{\mathcal{A}}
\def\skewprod#1#2{\langle #1 | #2 \rangle_{S} }
\def\MM{\left (M^{-1}\right )}
\def\Qdet#1{\mbox{Qdet} \left[ #1 \right]}
\def\Pf#1{\mbox{Pf}\ #1}
\def\PF#1{\mbox{Pf}\ \left[#1\right]}

\begin{document}
\topmargin -1.4cm
\oddsidemargin -0.8cm
\evensidemargin -0.8cm
\title{\Large{{\bf 
Massive partition functions and complex eigenvalue correlations in 
Matrix Models with symplectic symmetry\\
}}}

\vspace{1.5cm}
\author{~\\{\sc G.~Akemann}$^1$ and {\sc F.~Basile}$^{1,2}$
\\~\\
$^1$Department of Mathematical Sciences \& BURSt Research Centre\\
School of Information Systems, Computing and Mathematics\\
Brunel University West London\\ 
Uxbridge UB8 3PH, United Kingdom
\\~\\
$^2$Dipartimento di Fisica dell'Universit\'a di Pisa and INFN\\
Via Buonarroti, 56127 Pisa, Italy}

\date{}
\maketitle
\vfill
\begin{abstract}
We compute all massive partition functions or characteristic polynomials and
their complex eigenvalue correlation functions for non-Hermitean extensions of
the symplectic and chiral symplectic ensemble of random matrices. Our results
are valid for general weight functions without degeneracies of the mass
parameters. The expressions we derive are given in terms of the Pfaffian of
skew orthogonal polynomials in the complex plane and their kernel. They
are much simpler than the corresponding expressions for symplectic matrix
models with real eigenvalues, and we explicitly show how to recover these in
the Hermitean limit. This explains the appearance of three different kernels
as quaternion matrix elements there in terms of derivatives of a single kernel 
here.

\end{abstract}
\vfill

\thispagestyle{empty}
\newpage

\renewcommand{\thefootnote}{\arabic{footnote}}
\setcounter{footnote}{0}

\sect{Introduction}\label{intro}

Random Matrix Models with complex eigenvalues have received much attention
recently. This is due to both the development of new techniques leading to a
wealth of new analytical results, as well as to many interesting new 
applications, and we refer to 
\cite{FS,Zabrodin} for recent reviews and references. 
On the technical side new results were obtained on 
orthogonal polynomials in the complex plane such as
Hermite \cite{PdF} and Laguerre \cite{James,A05} polynomials, 
the solution of complex
Matrix Models (MM) using Replicas \cite{EK03} and their relation to the Toda
hierarchy \cite{SV1}, as well as sigma-model and supersymmetry methods
\cite{SV1,AFV}. All 3 approaches lead to the same results when applicable, and 
the two former \cite{AOSV} and the two latter have \cite{SV2} 
been shown to be equivalent.  

Our main motivation is the application of complex matrix models to Quantum
Chromodynamics (QCD) 
and related theories in the presence of a quark chemical potential $\mu$. 
Analytical MM predictions can be compared to numerical simulations in 
Lattice gauge theory. 
Here gauge group $SU(2)$ or the adjoint representation have the virtue that 
the Dirac spectrum remains real even in the presence of a chemical potential,
and standard Monte-Carlo techniques can be applied. 
For $SU(2)$ the sign of the  Dirac operator determinant fluctuates and
one has to restrict oneself to an even number of flavours.

In \cite{A05} a MM was introduced that describes the symmetry class for
theories in the adjoint representation with chemical potential. 
It is given by a two-MM 
generalising the chiral Symplectic Ensemble (chSE) \cite{NF1}
and was solved for two-fold degenerate quark masses \cite{A05}. 
Its predictions were
compared to quenched \cite{ABMLP} and unquenched Lattice simulations
\cite{AB}\footnote{In fact all simulations were unquenched using the code of
  \cite{HKLM}. The closest eigenvalues to the origin were then quenched
  (unquenched) using large (small) masses respectively.} for 2-colour QCD
using staggered fermions. 
Because of this success we expect that this MM is equivalent to the
corresponding chiral Lagrangian in the epsilon-regime \cite{Kogut}. 
In order to be able to prove this conjecture we complete here the solution of 
\cite{A05} by allowing for arbitrary non-degenerate quark masses. This will
permit a detailed comparison to group integrals of the
chiral Lagrangian in the epsilon-regime in the future. 
We also compute all massive partition functions (or
characteristic polynomials) and complex eigenvalue correlations
functions with arbitrary mass insertions  for another symmetry class, 
the complex extension of the Symplectic Ensemble (SE) \cite{Ginibre65}
for general weight functions  \cite{EK01}.

The expressions we obtain for the
complex SE and chSE are very similar in structure to the known results for the 
complex Unitary (UE) and chiral Unitary Ensemble (chUE), replacing
determinants by Pfaffian's. 
However, our computations are more difficult
as we have to use skew orthogonal instead of
orthogonal polynomials in the complex
plane. This is due to the Jacobian of symplectic matrices
\cite{Ginibre65,A05}. 
The difficulty is reflected in the fact that we can only compute products of
characteristic polynomials, extending \cite{AV}. It is not clear how to extend
the results for ratios in the unitary case \cite{Bergere,AP} to symplectic
ensembles. The proof for ratios in the SE with 
real eigenvalues \cite{BS} is
based on a discretisation of the ensemble
which is not obvious to extend to the complex case.

Our second main result concerns the link between symplectic MM with real and
complex eigenvalues. We use the fact that there exist weight functions in the
complex plane that
allow to take the Hermitean limit leading to real eigenvalue correlations,
of the SE and chSE respectively. In this way we explicitly 
recover the results \cite{Mehta,NF,NN,AK,NN2} for real eigenvalue 
correlation functions 
and characteristic polynomials for an arbitrary weight function. 
Moreover, from 
the Hermitean limit of our simple result for complex eigenvalue correlation
functions in terms of a Pfaffian of a single kernel 
we can derive why the known results for the 
chSE and SE are written in terms of a quaternion matrix of 
{\it three different} kernels, that are
related by differentiation. This follows very naturally from a Taylor
expansion of our single kernel in the Hermitean limit, due to the resulting
degeneracies of the variables $z$ with their complex conjugate $z^*$.

Our paper is organised as follows.
In the next section \ref{results} we give the definitions of our MM 
and their complex eigenvalue correlations in subsection \ref{defs},
and present the new results we obtain for massive correlation functions in 
subsection \ref{resultscorr}. Our results in the Hermitean limit are
summarised in subsection \ref{hermsum}.
The derivation of our findings are given in section \ref{proofs} where we prove
two theorems. Here, the chiral and non-chiral case are treated in parallel. 
At the end of this section we give examples for weight functions in either
case that allow to take the Hermitean limit.

\sect{Summary of Results}\label{results}

In this section we present our main 
results for complex extensions of the SE and
chSE to be defined below, with an arbitrary number of masses or characteristic
polynomials inserted. In the Hermitean limit we recover the known results 
for the SE and chSE with real eigenvalues.
We also briefly compare these findings to 
known results for 
the unitary complex MM.

\subsection{Definitions}\label{defs}

The massive partition function of the complex extension of the 
SE is defined as 
\be
{\cal Z}_N^{(M)}(\{m\})
\sim
 \int d\Phi  \prod_{f=1}^{M}\det[m_f\mathbb{\one}_N-\Phi]
\exp[-\Tr N V(\Phi,\Phi^\dag)] \ ,
\label{Zmatrix}
\ee
where $\Phi$ is an $N\times N$ matrix with quaternion real elements without
further symmetries, and $\mathbb{\one}_N$ is the quaternion unity element. 
The integral runs over all independent matrix
elements $\Phi_{ij}$. 
The $M$ mass parameters or arguments of the characteristic polynomials 
$m_f$ are taken to be 
complex and pairwise distinct, $m_i\neq m_j$. 

To proceed we only consider harmonic potentials of
the form $V(\Phi,\Phi^\dag)=\Phi\cdot\Phi^\dag+V_1(\Phi)+V_1(\Phi^\dag)$. 
The reason is that this choice in eq. (\ref{Zmatrix}) allows
to go to a complex eigenvalue basis by a Schur decomposition \cite{Mehta}
$\Phi=U(Z+T)U^\dag$. Here the diagonal matrix $Z$ with 
quaternion matrix element $Z_{ii}=$diag$(z_i,z_i^*)$ as a complex 
$2\times 2$ matrix contains the complex
eigenvalues in complex conjugate pairs. 
The upper triangular matrix $T$ drops out in the potentials $V_1$
and decouples in $\Tr\Phi\Phi^\dag=\Tr(ZZ^\dag+TT^\dag)$. 
Thus it can be integrated out being Gaussian, 
as well as the symplectic matrix $U$. For a detailed discussion of harmonic
potentials we refer to \cite{Zabrodin}.
After these manipulations we arrive at a complex eigenvalue integral given by
\be
{\cal Z}_N^{(M)}(\{m\})
\equiv
\int\prod_{i=1}^N d^2z_i\ w(z_i,z_i^*)  \prod_{f=1}^{M}(m_f-z_i)(m_f-z_i^*)
\ {\cal J}(\{z,z^*\})\ ,
\label{Zev}
\ee
with Jacobian \cite{Ginibre65}
\be
 {\cal J}(\{z,z^*\})\equiv
\prod_{k>l}^N |z_k-z_l|^2\ |z_k-z_l^*|^2
\prod_{h=1}^N |z_h-z_h^*|^2 \ .
\label{J}
\ee
Here $d^2z=dxdy$ with $z=x+iy$ denotes an integration over the full complex
plane.  In all the following we will take eq. (\ref{Zev}) as a starting
point. Furthermore we will only require from the weight function 
$w(z,z^*)=\exp[-V(z,z^*)]$ that i) it is real and symmetric
$w(z,z^*)=w(z^*,z)$ and that ii) all complex moments exist: 
$\int d^2z w(z,z^*) z^kz^{*\,l}<\infty$, being more general than in eq. 
(\ref{Zmatrix}). We note that the 
Jacobian considerably differs from the one of the SE given by the 
Vandermonde determinant $\Delta_N(\{x\})
\equiv\prod_{k>l}^N(x_k-x_l)$ of real eigenvalues
to the $4th$ 
power, ${\cal J}_{SE}=\Delta_N(\{x\})^4$. A complex extension of the SE 
with a similar Jacobian can be constructed from normal matrices, 
${\cal J}_{norm}=|\Delta_N(\{z\})|^4$ \cite{Hastings}. However, we have not
been able to obtain results for correlation functions 
in that case, even without mass insertions this
model is currently unsolved at finite-$N$ (for the spectral correlations from
a saddle point approximation at $N\to\infty$ see \cite{Zabrodin}).

Similar to eq. (\ref{Zev}) we define the complex extension of the chiral SE as 
\be
{\cal Z}_{N\,ch}^{(M)}(\{m\})
\equiv
\int\prod_{i=1}^N d^2z_i\ w(z_i,z_i^*)  \prod_{f=1}^{M}m_f^{2\nu}
(m_f^2-z_i^2)(m_f^2-z_i^{*\,2})\ {\cal J}_{ch}(\{z,z^*\})\ ,
\label{Zevch}
\ee
with Jacobian \cite{A05}
\be
 {\cal J}_{ch}(\{z,z^*\})\equiv
\prod_{k>l}^N |z_k^2-z_l^2|^2\ |z_k^2-z_l^{*\,2}|^2
\prod_{h=1}^N |z_h^2-z_h^{*\,2}|^2\ .
\label{Jch}
\ee
Note that in applications to QCD the mass terms are usually taken to be
positive, shifting $m\to im$. 
Eq. (\ref{Zevch}) 
only differs from the non-chiral ensemble by replacing its eigenvalues by
squares $z\to z^2$. 
While the complex SE has a repulsion of eigenvalues from
the real axis \cite{EK01}, as can be seen from the terms 
$|z_i-z_i^*|^2=4y_i^2$ 
in the Jacobian eq. (\ref{J}), the
occurrence of squared eigenvalues here leads to a repulsion from both the real
and imaginary axis \cite{A05}:  $|z_i^2-z_i^{*\,2}|^2=16x_i^2y_i^2$.
In addition we have added a factor $ \prod_{f=1}^{M}m_f^{2\nu}$,
ensuring a finite limit when $m_f\to0$.

The complex chSE enjoys a representation as a Gaussian two-matrix model 
of rectangular $N\times(N+\nu)$ matrices with real quaternion entries
\cite{A05}.
In this case the transformation to eigenvalues can be carried out
explicitly, and we refer to \cite{A05} for details. The outcome is that the 
resulting weight $w$ factorises into 
two parts: $w_V$ which results from inserting the
eigenvalue matrix $Z$ into the harmonic potential. This part is
non-universal. The second part $w_U$ comes about as follows. In the two-MM we
initially have 2 sets of complex eigenvalues. Since we are only interested in
the Dirac operator eigenvalues given by their product we change variables and
integrate out one set. The resulting part $w_U$ is not the exponential of the
potential, for an explicit example see eq. (\ref{wchGSE}). This factor 
is expected to be
universal as for the unitary ensembles it relates to bosonic partition
functions \cite{SV1,AOSV} (for a detailed discussion see \cite{SV3}).

In the following we will allow for a general weight $w$ in eq. (\ref{Zevch})
in terms of complex eigenvalues, with the only requirement of convergent 
moments as before. 
Expectation values of characteristic polynomials, which are proportional to
massive partition functions, are defined as 
\be
\left\langle \prod_{j=1}^N  \prod_{f=1}^M
(m_f-z_j)(m_f-z_j^*)\right\rangle_{{\cal Z}_N^{(0)}}\ \equiv\ 
\frac{{\cal Z}_N^{(M)}(\{m\})}{{\cal Z}_N^{(0)}} \ ,
\ee
and similarly for the chiral ensemble in terms of squared variables,
$m_f\to m_f^2$, $z_j^{(*)}\to z_j^{(*)2}$. The characteristic polynomials
also enjoy a matrix representation 
$\langle \prod_{f=1}^M \det[m_f\mathbb{\one}-\Phi]\rangle_{{\cal Z}_N^{(0)}}$
as in eq. (\ref{Zmatrix}).
We define the $k$-point complex
eigenvalues correlation functions in the presence of $M$ masses as 
\be
R_{N,k}^{(M)}(z_1,\ldots,z_k;\{m\})\equiv\frac{N!}{(N-k)!} 
\frac{1}{{\cal Z}_N^{(M)}(\{m\})}
\int\!\prod_{j=k+1}^N \!\!  d^2z_j
\prod_{i=1}^N w(z_i,z_i^*)\prod_{f=1}^{M}(m_f-z_i)(m_f-z_i^*)
{\cal J}(\{z,z^*\})
\label{defR_k}
\ee
where we integrate out all eigenvalues $z_l$ with $l\geq(k+1)$.
Obviously the  $k$-point function also depends on the complex conjugate
arguments $z_1^*,\ldots,z_k^*$ which we have suppressed in the notation, 
but not on the complex conjugate masses. 
In the chiral
expression we again have squared variables.

\subsection{Results for correlation functions}\label{resultscorr}

We obtain the following new result for arbitrary characteristic polynomials
\be
\left\langle \prod_{j=1}^N  \prod_{l=1}^M
(m_l-z_j)(m_l-z_j^*)\right\rangle_{{\cal Z}_N^{(0)}}\ =\ 
\frac{(-)^{[M/2]}}{\Delta_M(\{m\})}\Pf_{i,j=1,\ldots,M}
[\Theta_{N+[M/2]}(\{m\})]\ . 
\label{char}
\ee
Here we have to distinguish between even and odd $M$
\begin{equation}
\Theta_{N+[M/2]}(\{m\}) \equiv \left\{
\begin{array}{ll}
\left(\kappa_{N+[M/2]}(m_i,m_j)\right)_{i,j=1,\dots,M}& \ \ \mbox{ if $M$ is
  even} \\  
\left(\begin{array}{cc}
\kappa_{N+[M/2]}(m_i,m_j)_{i,j=1,\dots,M} & q_{2N+M-1}(m_i) \\
-q_{2N+M-1}(m_j) & 0
\end{array} \right)&\ \ \mbox{ if $ M $ is odd,}
\end{array} \right.
\label{theta}
\end{equation}
where $[M/2]$ denotes the integer part of $M/2$.
Thus for $M$ odd the $M\times M$ matrix $\Theta$ has 1 extra last row and
column. 
The polynomials $q_k(m)$ are the skew-orthogonal polynomials 
with respect to the following antisymmetric scalar product
\be
\langle f,g\rangle_S\ \equiv\ \int d^2z\ w(z,z^*)
(z^*-z)[f(z)g(z)^*-f(z)^* g(z)].
\label{scalar}
\ee
In the chiral case we simply modify the factor $(z^*-z)\to(z^{*\,2}-z^2)$ in
the scalar product.
They satisfy
\bea
\langle q_{2k+1},q_{2l}\rangle_S &=& -\langle q_{2l},q_{2k+1}\rangle_S
\ =\ r_k\ \delta_{kl}\nn\ ,\\
\langle q_{2k+1},q_{2l+1}\rangle_S &=& \, \ \ \langle q_{2l},q_{2k}\rangle_S \
\ \ \ =\ 0\ .
\label{skewdef}
\eea
In all the following we will chose them in monic normalisation,
$q_k(z)=z^k+{\cal O}(z^{k-1})$.
From these polynomials the second ingredient in 
eq. (\ref{theta}) is constructed, the anti-symmetric kernel\footnote{This is
  usually called pre-kernel as it does not include the weight function.}
\be
\kappa_N(z_1,z_2^\ast)\ \equiv\  
\sum_{k=0}^{N-1} \frac{1}{r_k} \left(q_{2k+1}(z_1)q_{2k}(z_2^\ast)-
q_{2k+1}(z_2^\ast)q_{2k}(z_1)\right)  \ .
\label{prekernel}
\ee
If we multiply eq. (\ref{char}) by the normalisation 
\be
{\cal  Z}_N^{(0)}=N!\prod_{i=0}^{N-1}r_i
\label{Znorm}
\ee 
we obtain the massive partition
functions  ${\cal Z}_N^{(M)}(\{m\})$.

Our second new result is for correlation functions with arbitrary  masses: 
\be
R_{N,k}^{(M)}(z_1,\dots, z_{k};\{m\})=\prod_{h=1}^k
w(z_h,z_h^\ast) (z_h^*-z_h) \frac{\Pf_{1,\dots,2k+M}
{\left[ \Theta_{N+[M/2]}(\{u\})
\right]}}{\Pf_{1,\dots,M}{\left[
      \Theta_{N+[M/2]}(\{m\})\right]}} \ ,
\label{Rkm}
\ee
where the set of variables in the numerator 
$\{u\}=\{z_1,z_1^\ast,\dots, z_k, z_k^\ast,
m_1,\dots,m_M\}$ runs through all masses, and all un-integrated complex
eigenvalues including their complex conjugates.
In general $\Theta$ is not the complex representation of a quaternion matrix.
The only modification of eq. (\ref{Rkm})
in the chiral case are squared arguments in the
Vandermonde in eq. (\ref{char}), $\Delta_M(\{m\})\to\Delta_M(\{m^2\})$
and in the prefactor in eq. (\ref{Rkm}), $(z_h^*-z_h)\to (z_h^{\,*2}-z_h^2)$.

In the case without masses $M=0$ eq. (\ref{Rkm}) reduces to the known result
\cite{EK01}, as we will show explicitly in the derivation in section
\ref{proofs} 
below. When $M$ is even and we choose the masses to appear in complex
conjugate pairs
we recover the results of \cite{A05}. In this case 
eqs. (\ref{char}) and (\ref{Rkm}) can be expressed entirely in terms of
$k$-point eigenvalue correlation functions without mass insertions
$R_{N,k}^{(0)}$ with $k=M$ 
and $k+M$ respectively \cite{A05}\footnote{For this relation to be exact we
  have to have an $N$-independent weight function.}. 

Let us give some examples. In the simplest case of a single mass or
characteristic polynomial we have 
\be
\left\langle \prod_{j=1}^N  
(m-z_j)(m-z_j^*)\right\rangle_{{\cal Z}_N^{(0)}}\ =\ 
q_{2N}(m)\ ,
\label{charex1}
\ee
giving the subset of even skew-orthogonal
polynomials in monic normalisation. 
This relation was already noted in \cite{EK01} in the complex SE
and in \cite{A05} for the complex chSE, and we can also write it as an
expectation value of a determinant as in eq. (\ref{Zmatrix}). 
A similar relation holds for the
odd skew-orthogonal polynomials as the expectation value of a single
determinant times the trace $\Tr(m\mathbb{\one}_N+\Phi)$ 
\cite{EK01,A05} (up to a constant), see
eq. (\ref{qodd}) below. Note that in
the general expression eq. (\ref{char}) only the even skew-orthogonal
polynomials appear explicitly (for $M$ odd) 
while the odd ones only occur through the
kernel. 

The second simplest example contains two characteristic polynomials,
\be
\left\langle \prod_{j=1}^N  
(m_1-z_j)(m_1-z_j^*)(m_2-z_j)(m_2-z_j^*)\right\rangle_{{\cal Z}_N^{(0)}}\ =\ 
\frac{r_N}{m_2-m_1}\ \kappa_{N+1}(m_2,m_1)\ .
\label{charex2}
\ee
This new relation gives the anti-symmetric kernel, the second building
block to all characteristic polynomials or eigenvalue correlation functions.
From eqs. (\ref{charex1}) and (\ref{charex2}) we could thus rewrite our
general results eqs. (\ref{char}) and (\ref{Rkm}) entirely in terms of
expectations values of one and two determinants. A similar structure has 
been revealed for MM with real eigenvalues in all 3 classical Wigner-Dyson
ensembles \cite{BS} (see also refs. in \cite{BDS}).

It is very instructive to compare eqs. (\ref{charex1}) and (\ref{charex2}) to
the known results  \cite{AV} for the complex extension of the UE and chUE:
both equations 
hold almost identically with the modification that we obtain {\it all}
orthogonal polynomials $P_k(z)$ with respect to 
$\int d^2z\ w(z,z^*) P_k(z)P_l(z^{*})=h_k\delta_{kl}$,
for both even and odd $k$  \cite{AV}: 
\be
\langle\det[m-\psi]\rangle_{UE}=P_N(m)\ .
\ee
Here $\psi$ is a complex non-Hermitean matrix and the polynomials 
are in monic normalisation $P_k(z)=z^k+{\cal O}(z^{k-1})$. The Hermitean 
kernel of the polynomials $P_k(z)$ exactly equals 
the expectation value of
a determinant times its complex conjugate  \cite{AV}:
\be
\langle\det[m_1-\psi]\det[m_2^*-\psi^\dag]\rangle_{UE}
=h_N\ K_{N+1}(m_1,m_2^*)\ ,
\ee 
where
$K_{N+1}(m_1,m_2^*)=\sum_{k=0}^Nh_k^{-1}P_k(m_1)P_k(m_2)$. In order to
underline the similarity in structure between 
the results for the complex unitary and symplectic ensembles we give the
result corresponding to eqs. (\ref{char}) and (\ref{Rkm}) as well. 
For products of characteristic polynomials of the UE we have \cite{AV}
\be
\left\langle \prod_{i=1}^K\det[m_i-\psi]\prod_{j=1}^L\det[n_j^*-\psi^\dag]
\right\rangle_{\!UE}\!\!\!\!=
\frac{\prod_{j=N}^{N+L-1}h_j}{\Delta_K(\{m\})\Delta_L(\{n^*\})}\\
\det_{1\leq i,j
\leq K+L}\left[
K_{N+L}(m_i,n_j^*) \ldots P_{N+L-j}(m_i)
\right] 
\label{charUE}
\ee
where $K\geq L$  without loss of generality.
For $K=L$ this is a determinant only made of kernels, 
as in eq. (\ref{char}) for $M$ even. 
In the symplectic ensemble the pairing of complex
conjugated eigenvalues is automatic, and  eq. (\ref{char}) resembles the square
root of eq.  (\ref{charUE}) at $m_i=n_i$. For $K>L$ there are $K-L$ extra
polynomials inside the determinant. 
For the massive eigenvalue correlations in the complex UE we have \cite{A01}
\be
R_{N,k\ UE}^{(M)}(z_1,\dots, z_{k};\{m\})=\prod_{h=1}^k
w(z_h,z_h^\ast) (z_h^*-z_h) \frac{\det_{1,\dots,2k+M}
{\left[ K_{N}(u_i,u_j)\right]}}{\det_{1,\dots,M}{\left[ 
K_{N}(m_i,m_j)\right]}} \ .
\label{RkmUE}
\ee
Its structure 
completely agrees with eq. (\ref{Rkm}) upon replacing the Pfaffian with a
determinant.

\subsection{Results in the Hermitean limit}\label{hermsum}

Suppose the weight function in the complex plane $w(z,z^*)$
depends on a non-Hermiticity parameter $\tau$ and permits a Hermitean limit
$\tau\to1$
that projects $z=x+iy$ onto the real $x$-axis:
\be
\lim_{\tau\to1} |z^*-z|^2 w(z,z^*) = \delta(y)\, \ow (x) \ ,
\label{hermlim}
\ee
where $\delta(y)$ is the Dirac delta-distribution\footnote{In \cite{A05} the
  limit was denoted by $-y\delta(y)'$ instead, which is equivalent after
  integration by parts.}. 
The main difference to the UE is that at any value $\tau<1$ the left hand
side appearing inside the integrals is identically zero for $y=0$,
the signature of the symplectic ensembles.
The weight $\ow (x)$ is the
projected weight function on $\mathbb{R}$, and we denote the 
projected skew orthogonal polynomials by $\oq_j(x)$, their norms by $\oer_j$ 
and their 
kernel by $\ok_N(x_1,x_2)$,
respectively. 

An example for such a weight function is the Gaussian weight of the complex SE
\bea
w_{GSE}(z,z^*)&=&\frac{
N^{\frac32}
\exp{\left( -\frac{N}{1-\tau^2}\left(|z|^2-\frac{\tau}{2}(z^2+z^{\ast
    2})\right)\right)}}{2\sqrt{\pi}(1-\tau)^{\frac32}} 
\nn\\
&=&\frac{N^{\frac32}}{2
\sqrt{\pi}(1-\tau)^{\frac32}}\exp{\left(-\frac{Ny^2}{1-\tau}\right)} 
\exp{\left(-\frac{Nx^2}{1+\tau}\right)}\ , \ \ \ \tau\in[0,1)\ ,
\label{wGSE}
\eea
with a resulting projected Gaussian weight $\ow (x)=\exp(-Nx^2/2)$.
Before the Hermitean limit $\tau \to 1$ is taken 
the corresponding 
skew orthogonal polynomials are given in terms of Hermite polynomials
in the complex plane \cite{EK01}. After the projection they are given by the
ordinary skew orthogonal polynomials 
of the Gaussian SE in terms of Hermite polynomials on the real line 
\cite{Mehta}. 
A similar example exists for the chiral
ensemble in terms of Laguerre polynomials
in the complex plane \cite{A05}, and we refer to subsection \ref{Th2} for more
details.   
The important point is that the Hermitean limit maps our previous results eqs. 
(\ref{char}) and (\ref{Rkm}) to the known results of the symplectic ensembles
with real eigenvalues. 

More explicitly 
the formula (\ref{char}) for the characteristic polynomials we trivially
replace the polynomials and kernel by their projected quantities inside the
matrix $\Theta$ in eq. (\ref{theta}):
\be
\lim_{\tau\to1}
\left\langle \prod_{j=1}^N  \prod_{l=1}^M
(m_l-z_j)(m_l-z_j^*)\right\rangle_{{\cal Z}_N^{(0)}}\ =\ 
\frac{(-)^{[M/2]}}{\Delta_M(\{m\})}\Pf_{1,\ldots,M}[\overline 
\Theta_{N+[M/2]}(\{m\})]\  
\label{charH}
\ee
The limit for correlation functions is more involved, and we obtain the
following non-trivial result:
\be
\lim_{\tau\to1}R_{N,k}^{(M)}(z_1,\dots, z_{k};\{m\})=\prod_{h=1}^k
\delta(y_h)
\overline w(x_h) \frac{\Pf_{1,\ldots,2k+M}{\left[
      \OO_{N+[M/2]}(x_1,x_2,\dots,x_k,m_1,\dots,m_M) 
\right]}}{\Pf_{1,\ldots,M}{\left[
      \OO_{N+[M/2]}(\{m\})\right]}} \ .
\label{RkmH}
\ee
To compare to real eigenvalue correlation functions we have to drop the
delta-functions, or formally integrate over the remaining imaginary parts 
$y_1,\dots,y_k$. The  projected matrix  $\OO_{N+[M/2]}$
is obtained as 
\begin{equation}
\overline \Omega_R\equiv
\left(
\begin{array}{ccccccc}
\dots & \dots & \dots & \dots & \dots & \dots & \dots\\
\dots & \partial_{x_i} \partial_{x_j}\overline \kappa_R(x_i,x_j) &
\partial_{x_i} \overline \kappa_R(x_i,x_j) & \dots & \partial_{x_i}\overline
\kappa_R(x_i,m_f) &\dots & \partial_{x_i} \overline q_{2R}(x_i)\\ 
\dots &\partial_{x_j}\overline \kappa_R(x_i,x_j) & \overline\kappa_R(x_i,x_j) &
\dots & \overline\kappa_R(x_i,m_f) &\dots & \overline q_{2R}(x_i)\\ 
\dots & \dots & \dots & \dots & \dots & \dots\\
\dots & \partial_{x_j} \overline \kappa_R(m_g, x_j) &
\overline\kappa_R(m_g,x_j) & \dots & \overline \kappa_R(m_g,m_f)
&\dots & \overline q_{2R}(m_g) \\ 
\dots & \dots & \dots & \dots & \dots & \dots & \dots \\
\dots & - \partial_{x_j} \overline q_{2R}(x_j) & -\overline
q_{2R}(x_j) & \dots & -\overline q_{2R}(m_f) & \dots & \dots\\ 
\end{array}
\right)
\label{Om}
\end{equation}
and depends on the projected kernel and its first and second derivatives. 
Here $R=N+[M/2]$ and $2R=2k+M-1$. If we now identify 
\bea
I_N(x,t)&=&
\sqrt{\overline w(x) \overline w(t)}\ 
\overline \kappa_N (x,t)\nn \\
S_N(x,t)&=&
\sqrt{\overline w(x) \overline w(t)}\  
\partial_t \overline \kappa_N(t,x) 
\ =\ -\sqrt{\overline w(x) \overline w(t)}\  
\partial_t \overline \kappa_N(x,t) 
\nn\\ 
D_N(x,t)&=&
-\sqrt{\overline w(x) \overline w(t)}\ 
\partial_x \partial_t \overline\kappa_N (x,t) 
\label{kappaISD}
\eea
with the standard notation for the matrix elements of the quaternion valued
kernel \cite{Mehta,NN2} 
we recover the following known results for symplectic matrix models on the
real line:
\begin{itemize}

\item[i)] the result eqs. (\ref{charH}) was derived in \cite{BS} for the
  Gaussian SE
  in terms of expectation values of two characteristic polynomials, or
  equivalently in terms of the $I_N(x,t)$-kernel \cite{NN2}. 
  In taking the Hermitean limit we now 
  understand why only the $I$- and not the $S$- and $D$-kernels
  appear. Moreover, the result from \cite{BS} is strictly valid only for a
  Gaussian weight function, whereas we can allow for an arbitrary weight here
  (for details see the derivation below)\footnote{In 
  \cite{NN} the authors restrict themselves to a Gaussian weight. A closer
  analysis of the derivation that follows from an earlier paper \cite{NF}
  shows that this restriction can be lifted.}. For the chiral ensembles we
  recover the results of \cite{NN,AK}.

\item[ii)] In eq. (\ref{RkmH}) we recover 
the well known result
  \cite{Mehta} for correlation functions of the Gaussian SE in the absence of
  masses, which are given equivalently in terms of a quaternion determinant. 
We also re-obtain the massive correlation functions
  from \cite{NN2}, and in the chiral case the correlation functions 
\cite{NF1} and  \cite{NN,AK}, respectively. Before taking the Hermitean limit 
our results are not only simpler, but they also offers an explanation why
  in the Hermitean limit three different kernels appear that are the first and
  second derivative of the kernel 
$\overline{\kappa}_N(x,t)$. 
\end{itemize}

The fact that the 3 kernels in eq. (\ref{kappaISD})
    are related by differentiation was of course known previously, but without
  a deeper reasoning behind.
So why does the complexification $x_i\to z_i$ 
simplify? The answer is that it lifts the following degeneracy: 
when replacing the
Jacobian for real eigenvalues $\Delta(\{x\})^4$ by a determinant one has to
use polynomials and their derivatives \cite{Mehta}. 
Our Jacobians eqs. (\ref{J}) and  (\ref{Jch}) are proportional to a single
Vandermonde determinant and can thus be expressed in terms of polynomials
only, see eq. (\ref{DeltaP}) below.
Furthermore we are able to build a full $2\times2$ matrix depending on a
single kernel of two arguments $u_i$ and $u_j$ and their complex conjugates. 
For real eigenvalues this is not possible.

As a final remark we comment on the other ensembles. 
For the Hermitian limit of the complex 
UE and chUE some interesting identities follow, and we refer to \cite{AV}. 
On the other hand for complex extensions of the orthogonal ensemble 
in terms of real non-symmetric matrices 
we do not expect any simplification to happen. 
To date no closed formula is known for their correlation functions, 
even without mass insertions.


\sect{Derivation of Results}\label{proofs}

\subsection{Complex eigenvalue integrals}

In this subsection we prove the following theorem in terms of a
Pfaffian for $(N-k)$-fold 
integrals over complex eigenvalues with $M$ mass insertions.

\noi
{\sc Theorem 1} 
{\it Let $ z_i\ \in\ \mathbb C, i=1,\ldots N $ be complex variables 
and $m_f \ \in\ \mathbb C, f=1,\ldots M $ be complex mass parameters. 
Given a real weight function $w(z,z^*)$ 
defined in the whole complex plane such that all moments exits, 
$\int d^2z\ w(z,z^*) z^kz^{*\,l}<\infty$, and a set of skew orthogonal
polynomials $q_k(z)$ satisfying eq. (\ref{skewdef}) with scalar product
(\ref{scalar}). Then the following integral can be computed as} 
\bea
 \aa_{k,N}^{(M)}
(z_1, \dots, z_k;m_1,\dots, m_M)&\equiv& \int \prod_{j=k+1}^N d^2z_j \
\prod_{i=1}^N w(z_i,z_i^*)  \prod_{f=1}^{M}(m_f-z_i)(m_f-z_i^*)
{\cal J}(\{z,z^*\})
\nn
\\
&=& \frac{(-)^{[M/2]}(N-k)!}{\Delta_M(\{m\})} 
\prod_{h=1}^k w(z_h,z_h^\ast)  (z^{\ast}_h-z_h) 
\!\! 
\prod_{h=0}^{N+[M/2]-1}\!\!r_h\nn\\
&&\times\Pf_{1,\dots,2k+M}[\Theta_{N+[M/2]}(\{u\})]\ ,
\label{Afunction}
\eea
{\it 
with the set $\{u\}=\{z_1,z_1^\ast,\dots, z_k, z_k^\ast,
m_1,\dots,m_M\}$, 
${\cal J}(\{z,z^*\})$ defined by eq. (\ref{J}) and the 
matrix $\Theta$ defined by 
eqs. (\ref{theta}) and (\ref{prekernel}) respectively.
For the corresponding chiral integral with Jacobian eq. (\ref{Jch}) we obtain}
\bea 
\aa_{k,N\,ch}^{(M)}
(z_1, \dots, z_k;m_1,\dots, m_M)&\equiv& \int \prod_{j=k+1}^N d^2z_j \
\prod_{i=1}^N w(z_i,z_i^*)
 \prod_{f=1}^{M}
(m_f^2-z_i^2)(m_f^2-z_i^{*\,2}) 
{\cal J}_{ch}(\{z,z^*\})\nn\\
&=& 
\frac{(-)^{[M/2]}(N-k)!}{\Delta_M(\{m^2\})} 
\prod_{h=1}^k w(z_h,z_h^\ast)  (z^{\ast\,2}_h-z_h^2) 
\!\! \prod_{h=0}^{N+[M/2]-1}\!\!r_h\nn\\
&&\times\Pf_{1,\dots,2k+M}[\Theta_{N+[M/2]}(\{u^2\})]\ ,
\label{Afunctionch}
\eea
{\it 
using the skew orthogonal polynomials $q_k(z)$ and kernel 
of the corresponding chiral 
scalar product.  }

\indent
 
Note that the integrals $\aa$ defined above depend also on the complex
conjugated eigenvalues $z_1^*,\dots,z_k^*$, but not on the conjugated masses.

\noi
{\sc Proof}:
The main ingredient of the proof is the
availability of a skew orthogonal basis. 
For simplicity we will give the proof only for the non-chiral
case, eq. (\ref{Afunction}). 
The chiral case trivially follows along the very same lines, inserting squared
variables and using the chiral skew orthogonal product.

The first step is to write the integrand of (\ref{Afunction}) 
in terms of a Vandermonde determinant. For this aim we explicitly write out
the Jacobian eq. (\ref{J}) 
\bea
&&\prod_{i=1}^N \prod_{f=1}^M (m_f-z_i)(m_f-z_i^*)
\prod_{k>l}^N |z_k-z_l|^2 |z_k-z_l^{\ast}|^2
\prod_{j=1}^N |z_j - z_j^{\ast}|^2 = \nn\\
&&=(-)^N\prod_{k>l}^N (m_k- m_l)^{-1} \prod_{j=1}^N (z^{\ast}_j-z_j) 
\Delta_{2N+M}(z_1, z_1^{\ast},\dots, z_N, z_N^{\ast}, m_1, \dots, m_M)\ ,
\eea
where we have used that the masses $m_f$ are pairwise distinct.
Using the standard trick a Vandermonde determinant 
may be written in terms of any set of monic polynomials 
$ p_{j-1}(v_i) $ of degree $ j-1 $:
\be
\Delta_n(v_1,\dots,v_n) =\det_{i,j=1,\dots,n}[v_i^{j-1}]
=\det_{i,j=1,\dots,n}[p_{j-1}(v_i)]
\label{DeltaP}
\ee
Here we chose the set of complex skew orthogonal polynomials $ q_i(z)$
that satisfies the skew orthogonality eq. (\ref{skewdef}) with respect to our
weight function.
Defining the set of $2N+M$ variables 
$\{v_1,\dots,v_{2N+M}\}=\{z_1,z_1^\ast,\dots,z_{N},z_{N}^*,m_1,\dots,m_M\}$ 
we can write (suppressing the arguments of $\aa$)
\bea
\aa_{k,N}^{(M)}&=&\frac{(-)^N}{\Delta_M(\{m\})}
\int\prod_{h=k+1}^N d^2z_h \prod_{j=1}^N w(z_j,z_j^\ast)\ (z_j^{\ast}-z_j) \
\det_{i,j=1,\dots,2N+M} [q_{j-1}(v_i )] \\ 
&=&\frac{(-)^N}{\Delta_M(\{m\})}
\sum_{\{a\}}\sigma(a)\ \int\prod_{h=k+1}^N d^2z_h \prod_{j=1}^N
w(z_j,z_j^\ast)\ (z_j^{\ast}-z_j)\prod_{j=0}^{2N+M-1} q_{a_{j}}(v_{j+1}). 
\eea
The summation runs over all the possible permutations of 
$ \{0,1,\dots,2N+M-1 \} $ with sign $\sigma(a)$. 
The integrand is symmetric with respect to exchanging
$z_h\leftrightarrow z_h^*$ for all integrated variables
$z_{k+1},\ldots,z_{N}$. Without loss of
generality we can therefore 
arrange the sum over all permutations such that it always holds
$a_{2j}<a_{2j+1} \ \forall j: \ k\leq j<N$.
We denote this 
rearrangement as  $ \sum_{\{a\}'} $.
After this manipulation we can write
\bea
\aa_{k,N}^{(M)}&=&\frac{(-)^N}{\Delta_M(\{m\})}
\sum_{\{a\}'}
 \sigma(a) \prod_{h=1}^k w(z_h,z_h^\ast)  (z^{\ast}_h-z_h) 
\prod_{i=0}^{2k-1} q_{a_{i}}(v_{i+1}) 
\prod_{l=2N}^{2N+M-1 }   q_{a_{l}}(v_{l+1})  \nn\\
&&\times  
\int\prod_{j=k}^{N-1}d^2z_{j+1}w(z_{j+1},z_{j+1}^\ast)(z^{\ast}_{j+1}-z_{j+1}) 
(q_{a_{2j}}(z_{j+1})  q_{a_{2j+1}}(z^{\ast}_{j+1}) -
q_{a_{2j+1}}(z_{j+1})  q_{a_{2j}}(z^{\ast}_{j+1}) )\nn\\ 
&=& \frac{(-)^N}{\Delta_M(\{m\})}
\sum_{\{a\}'}\! \sigma(a)\prod_{h=1}^k w(z_h,z_h^\ast)(z^{\ast}_h-z_h) 
\prod_{i=0}^{2k-1} q_{a_{i}}(v_{i+1})\! \!
\prod_{l=2N}^{2N+M-1 }\!\!  q_{a_{l}}(v_{l+1}) 
\prod_{j=k}^{N-1} \skewprod{q_{a_{2j}}}{q_{a_{2j+1}}}\nn\\
&& \label{Afunction2}
\eea
taking out the un-integrated variables. 
Using the properties (\ref{skewdef}) we can see that those permutations
giving a non-zero value will be the ones satisfying: 
\begin{equation}
a_{2j}\ +\ 1=a_{2j+1} \ \ { and } \ \ a_{2j} \mbox{ is even} \ \forall\ j:
k\leq j<N \ .
\end{equation}
Hence for all other indices $ a_{l} $ with $l\notin\{2k+1,\dots,2N\} $ of the
non-integrated polynomials only the following configurations contribute. 
The $a_l$ are coupled in pairs as for even indices $a_l$ their successor $a_l\
+\ 1$ and for odd indices $a_l$ their predecessor $a_l\ -\ 1$ 
cannot belong to the integrated polynomials: 
\begin{itemize}
\item $ a_{l} \mbox{ is even}\ and\ a_l\neq 2N+M-1 \ \Rightarrow \exists\
  l^\prime\notin\{2k+1,\dots,2N\} $ such that $a_{l^\prime} = a_{l}\ +\ 1$  
\item $ a_l \mbox{ is odd }\ 
\Rightarrow \exists\ l^\prime\notin\{2k+1,\dots,2N\} $
  such that $ a_{l^\prime}\ +\ 1=a_l $ \ .
\end{itemize} 
There is only one possibility that $ a_l $ is un-coupled: 
\begin{itemize}
\item $ M $ is odd and $ a_l $ assumes the maximum value\footnote{We have
an odd number of variables here.}, $ a_l=2N+M-1 $ 
\end{itemize}
This coupling into pairs is thus 
invoked from the scalar products in the last term
in eq. (\ref{Afunction2}).

We rename the set of the $2k+M$ non-integrated 
variables $ \{u_j\}=\{ z_1,z_1^\ast, \dots,z_k, z_k^{\ast},m_1,\dots,m_{M}\}$. 
Eq. (\ref{Afunction2}) can be seen as the sum of configurations
anti-symmetrised 
with respect to these variables $ \{u\} $ - we call such permutations 
$ \eta $ with sign $\sigma(\eta)$ - 
and moving the indices of subsequent non-integrated polynomials 
$q_{2j}(\ )q_{2j+1}(\ ) $ along the line $ j=0,\dots,N+[M/2]$. 
In order to
count terms only once we must divide by $ L! $, with $ L\equiv k+[M/2] $, as
these are the number of possible permutations to put $L$ variables into pairs 
$q_{2j}(\ )q_{2j+1}(\ ) $.

In the next step we show how these sums can be written in terms of the
kernel eq. (\ref{prekernel}). First of all there are $(N-k)!$ permutations
of the arguments of the integrated variables inside the scalar products that
all give the same contribution. If we multiply and divide by all the remaining
norms $r_h$ we can write
\bea
\aa_{k,N}^{(M)}&=&\frac{(-)^N}{\Delta_M(\{m\})L!}
\prod_{h=1}^k w(z_h,z_h^\ast)  (z^{\ast}_h-z_h) (N-k)!
\prod_{h=0}^{N+[M/2]-1}(-r_h) 
\label{Afunction3}\\
&&\times \sum_{\{\eta\}} \sigma(\eta) \ 
\sum_{\stackrel{h_1,\dots,h_L=0}{h_l\neq h_j}}^{N+[M/2]-1}  
\prod_{i=1}^L \frac{1}{(-r_{h_i})} q_{2h_i}(u_{\eta_{2i-1}})
q_{2h_i+1}(u_{\eta_{2i}}) 
\left\{
\begin{matrix} 1 & \mbox{$M$ even} 
\\ q_{2N+M-1}(u_{\eta_{2L+1}}) & \mbox{$M$ odd}
\end{matrix} \right.   \nn\\
&=&\frac{(-)^{[M/2]}}{\Delta_M(\{m\})2^L L!}
\prod_{h=1}^k w(z_h,z_h^\ast)  (z^{\ast}_h-z_h)
(N-k)! \ \prod_{h=0}^{N+[M/2]-1}r_h  \nn\\ 
&&\times \sum_{\{\eta\}} \sigma(\eta) \ \prod_{i=1}^L
\kappa_{N+[M/2]}(u_{\eta_{2i-1}},u_{\eta_{2i}}) 
\left\{\begin{matrix} 1 & \mbox{ $M$ even} 
\\ q_{2N+M-1}(u_{\eta_{2L+1}}) & \mbox{ $M$ odd}
\end{matrix} \right. \ .\label{Afunction4}
\eea
To go from (\ref{Afunction3}) to (\ref{Afunction4}) we have used the 
antisymmetry to generate all the terms in the sum of the kernel, giving a
factor of $1/2^L$. Furthermore, we have used that where $ h_l=h_j $ the terms 
drop out due to the antisymmetry in the $ \{u\} $. 

In a final step we show 
that the second line 
of (\ref{Afunction4}) may be written in terms of a 
Pfaffian. For that we distinguish between even and odd $M$.

\paragraph{M even:}

We define the antisymmetric matrix of size $ 2L=2k+M$
\begin{equation} \label{kappa_even}
\Theta_{R}(\{u\})\equiv
\begin{pmatrix}
0 & \kappa_R(u_1,u_2) & \cdots & \kappa_R(u_1,u_{2L}) \\
\kappa_R(u_2,u_1) & 0 & \cdots & \kappa_R(u_2,u_{2L}) \\
\vdots & & \ddots & \vdots \\
\kappa_R(u_{2L},u_1) & \kappa_R(u_{2L},u_2) & \cdots & 0
\end{pmatrix}=
(\kappa_R(u_i,u_j))_{i,j=1\dots 2L}\ .
\end{equation}
For this matrix the Pfaffian written as an ordered expansion \cite{Mehta}
is given by 
\begin{equation}
\Pf_{i,j=1\dots 2L}[\kappa_R(u_i,u_j)]
=\frac{1}{2^L L!} \sum_{\{\eta\}} \sigma(\eta)
\prod_{j=1}^L \kappa_R(u_{\eta_{2j-1}},u_{\eta_{2j}})\ . 
\end{equation}
For $ R=N+[M/2] $ this is exactly the desired result.

\paragraph{M odd:} 

We define the antisymmetric matrix of size $ 2L+2=2k+M+1$ 
\bea
\label{kappa_odd}
\Theta_{R}(\{u\})&=&
\begin{pmatrix}
0 & \kappa_R(u_1,u_2) & \cdots & \kappa_R(u_1,u_{2L+1}) & q_{2R}(u_1)\\
\kappa_R(u_2,u_1) & 0 & \cdots & \kappa_R(u_2,u_{2L+1}) & q_{2R}(u_2)\\
\vdots & & \ddots & \vdots & \vdots\\
\kappa_R(u_{2L+1},u_1) & \kappa_R(u_{2L+1},u_2) & \cdots & 0 &
q_{2R}(u_{2L+1})  \\ 
-q_{2R}(u_1) & -q_{2R}(u_2) & \dots &-q_{2R}(u_{2L+1}) & 0
\end{pmatrix}\nn\\
&=&\begin{pmatrix}
(\kappa_R(u_i,u_j))_{i,j=1\dots 2L+1} & q_{2R}(u_i) \\
-q_{2R}(u_j) & 0
\end{pmatrix}\ .
\eea
Consider the Pfaffian of this matrix 
\begin{equation}
\label{pfaffian_odd}
\Pf_{i,j=1\dots 2L+2}[\Theta_R(\{u\})]=\frac{1}{2^{L+1}
  (L+1)!}\sum_{\eta}\sigma(\eta) \prod_{j=1}^{L+1}
\Theta_R(\{u\})_{\eta_{2j-1},\eta_{2j}}  \ .
\end{equation}
In every permutation there exists exactly one index $i$ such that 
$ \eta_i=2L+1 $. The idea is to manipulate the permutations $\eta$ over $2L+2$
elements to obtain a sum of permutations $\eta'$ over $2L$ elements, leading
to eq. (\ref{Afunction4}). 

As a first step we verify the invariance of the product 
$ \sigma(\eta)\prod_{j=1}^{L+1}
\Theta_R(\{u\})_{\eta_{2j-1},\eta_{2j}}$ under the following 
elementary transformations:
\begin{itemize}
\item[i)] exchange of indices of a single factor 
$\Theta_R(\{u\})_{\eta_{2j-1},\eta_{2j}}$: 
$ \eta_{2j-1} \longleftrightarrow \eta_{2j} $
\item[ii)] exchange of 2 pairs of indices among two factors:
$ \eta_{2j-1},\eta_{2j} \longleftrightarrow \eta_{2j+1},\eta_{2j+2}$
\end{itemize}
In the first transformation the change in sign is compensated by $\sigma(\eta)$
and in the second transformation 
the sign remains. Composing these two elementary steps we
can always achieve that the factor containing the index $ \eta_i=2L+2 $ is 
commuted  to the end:
\begin{itemize}
\item if $ i $ is even: 
$\sigma(\eta) \prod_{j=1}^{L+1}
\Theta_R(\{u\})_{\eta_{2j-1},\eta_{2j}}=\sigma(\eta) \prod_{j=1}^{L}
\Theta_R(\{u\})_{\eta_{2j-1},\eta_{2j}}\Theta_R(\{u\})_{\eta_{2j+1}
,\eta_i=2L+2}$  
\item if $ i $ is odd: 
$\sigma(\eta) \prod_{j=1}^{m+1}
\Theta_R(\{u\})_{\eta_{2j-1},\eta_{2j}}= - \sigma(\eta) \prod_{j=1}^{L}
\Theta_R(\{u\})_{\eta_{2j-1},\eta_{2j}}\Theta_R(\{u\})_{\eta_{2j+1}
,\eta_i=2L+2}$ .
\end{itemize}
In the first case we have used only ii), in the second we used both i) and
ii). In each case we can thus write the result as a permutation $\eta'$ of 
only $2L$ elements: 
\begin{equation}
\sigma(\eta) \prod_{j=1}^{L+1}
\Theta_R(\{u\})_{\eta_{2j-1},\eta_{2j}}=\sigma(\eta') \prod_{j=1}^{L}
\Theta_R(\{u\})_{\eta_{2j-1}',\eta_{2j}'}\Theta_R(\{u\})_{\eta_{2j+1}',2L+2}\ .
\label{equiv_perm}
\end{equation}
Here $2(L+1)$ permutations $\eta$ will be mapped to the same permutation 
$\eta'$: there are $(L+1)$ ways to choose the second index corresponding to 
$\eta_i$, and it can be in the first ($i$ even) or second place ($i$ odd) of
the index pair. Together we obtain from eq. (\ref{pfaffian_odd}) that
\begin{equation}
\Pf_{i,j=1\dots 2L+2}[\Theta_R(\{u\})]=\frac{1}{2^{L+1} (L+1)!}2(L+1) 
\sum_{\{\eta^\prime\}} \sigma(\eta^\prime) \prod_{j=1}^L
\kappa_R(y_{\eta^\prime_{2j-1}},y_{\eta^\prime_{2j}})
q_{2R}(y_{\eta^\prime_{2L+1}}) \ ,
\end{equation}
where $\eta'$ runs over $2L$ elements and we have inserted the explicit 
matrix elements from eq. (\ref{kappa_odd}).
That is the desired result for $ R=N+[M/2],\ 2R=2N+M-1 $.

To summarise we have shown
\begin{equation}
 \aa_{k,N}^{(M)}
(z_1, \dots, z_k;m_1,\dots, m_M)
= \frac{(N-k)!(-)^{[M/2]}}{\Delta_M(\{m\})}\prod_{h=1}^k w(z_h,w_h^\ast)
(z^{\ast}_h-z_h)  \!\!\!
\prod_{h=0}^{N+[M/2]-1}\!\!\!r_h\ \Pf[\Theta_{N+[M/2]}(\{u\})] 
\end{equation}
where $ \Theta_{N+[M/2]}(\{u\}) $ is defined in (\ref{kappa_even}) for even $
M $ and in (\ref{kappa_odd}) for odd $ M $, as announced in
eq. (\ref{theta}). $\Box$\\

\indent
Let us point out how all the results in subsection 
\ref{resultscorr} follow from Theorem 1.
For $k=M=0$ 
the Pfaffian and prefactors are absent and we obtain the
normalisation of the massless partition function eq. (\ref{Znorm})
\be
{\cal  Z}_N^{(0)}= \aa_{k=0,N}^{(M=0)}=N!\prod_{i=0}^{N-1}r_i \ .
\ee
For $k=0$ and $M>0$ we obtain the massive partition functions 
${\cal  Z}_N^{(M)}(\{m\})$
\be
{\cal  Z}_N^{(M)}(\{m\})= \aa_{k=0,N}^{(M)}(m_1,\dots, m_M)\ ,
\ee
and after dividing by ${\cal  Z}_N^{(0)}$ the characteristic polynomials 
eq. (\ref{char}). Finally the complex eigenvalues correlation functions are
obtained as 
\be
R_{N,k}^{(M)}(z_1,\dots, z_{k};\{m\})=\frac{N!}{(N-k)!} \frac{1}{{\cal
    Z}_N^{(M)}(\{m\})}  \aa_{k,N}^{(M)}
(z_1, \dots, z_k;m_1,\dots, m_M)\ .
\ee
The Vandermonde determinant of the mass parameters cancels as well as all
factorials and the product over norms $r_h$, leading to the ratio of Pfaffian 
in eq. (\ref{Rkm}). In the same way all the results for the chiral ensemble
follow, where for the partition functions we have to multiply in the factor 
$\prod_{f=1}^{M}m_f^{2\nu}$.

As a check we can recover the results of the non-chiral 
model without mass insertion \cite{EK01}\footnote{\label{footnote_sign}The 
result is the same up to an overall
  factor. The proof in \cite{EK01} follows the one in \cite{Tracy1998}, that is
  derived for the squared quantities, hence the result is true up to an
  overall sign.} expressed in term of quaternion determinants \cite{Dyson1972}:
\bea
R_{N,k}^{(0)}(z_1,\dots, z_{k})
&=&(-)^k\prod_{h=1}^k w(z_h,z_h^\ast) (z^{\ast }_h-z_h)\ 
\Qdet{\begin{pmatrix} \kappa_N(z_i^\ast,z_j) & -\kappa_N(z^\ast_i,z^\ast_j)
\\ \kappa_N(z_i,z_j) & -\kappa_N(z_i,z^\ast_j) \end{pmatrix}_{i,j=1\dots k} }
\nn\\ 
&=&(-)^k\prod_{h=1}^k w(z_h,z_h^\ast) (z^{\ast }_h-z_h)\ \Pf{\left[
    \begin{pmatrix} \kappa_N(z_i,z_j) & -\kappa_N(z_i,z_j^\ast) \\
      -\kappa_N(z_i^\ast,z_j) & \kappa_N(z_i^\ast,z_j^\ast)
    \end{pmatrix}_{i,j=1\dots k} \right]} \nn\\ 
&=&\prod_{h=1}^k w(z_h,z_h^\ast) (z^{\ast }_h-z_h)\ \Pf{\left[ \begin{pmatrix}
      \kappa_N(z_i,z_j) & \kappa_N(z_i,z_j^\ast) \\ \kappa_N(z_i^\ast,z_j) &
      \kappa_N(z_i^\ast,z_j^\ast) \end{pmatrix}_{i,j=1\dots k} \right]} \nn\\ 
&=&\prod_{h=1}^k w(z_h,z_h^\ast) (z^{\ast }_h-z_h) \
\Pf{\left[ \Theta_N(z_1,z_1^\ast,\dots, z_k,z_k^\ast)\right]}\ .
\label{rho_nc_as_qdet}
\eea
Here we have used two properties of the Pfaffian, one is the well known 
identity by Dyson \cite{Dyson1972}  
\begin{equation}
\Qdet{A}=\PF{C\left(\Sigma\otimes \mathbb{\one}_N\right)\cdot C(A)}  
\end{equation} 
where $A$ is an $N\times N$ quaternion matrix, 
$\Sigma=\begin{pmatrix} 0 & 1\\ -1 & 0\end{pmatrix}$ , and  
$C(A)$ is the $2N\times 2N$ complex representation of $A$.
The other identity 
is that for a matrix $B$ and an anti-symmetric matrix $A$, both
of the same size, the following holds \cite{Mehta2}  
\be
\mbox{Pf}[B\,A\,B^T]\ =\ \mbox{Pf}[A]\ \det[B]\ .
\label{ABid}
\ee
Using for $B=\begin{pmatrix} 1 & 0\\ 0& -1\end{pmatrix}\otimes \mathbb{\one}_k$
with determinant $\det[B]=(-)^k$
we can equate the second and third line above in eq. 
(\ref{rho_nc_as_qdet}). In the last step we used the
definition eq. (\ref{theta}).

It is relatively easy to see that for pairs of complex
conjugated masses the correlation functions in eq. (\ref{Rkm}) can be
expressed in terms of ratios of  correlation functions without mass insertions
\cite{A05}. 


\subsection{The Hermitean limit}\label{Th2}

\noi
{\sc Theorem 2 (Hermitean limit)}
{\it
Given the real weight function from Theorem 1 
is depending on a non-Hermiticity parameter $ \tau $, $w=w(z,z^\ast;\tau)$,
and the following limit exist for any finite number of
eigenvalues $N$
\begin{equation} 
\label{hypo_limit}
\lim_{\tau\to1} |z^*-z|^2 w(z,z^*) = \delta(y)\ \ow (x) \ .
\end{equation}
It is called the Hermitean limit and  $\ow (x)$ is the projected weight
function on $\mathbb{R}$.
It then follows that the Hermitean limit of the integrals computed in Theorem 1
exists and is given by} 
\begin{equation}
\lim_{\tau\to 1} 
\aa_{k,N}^{(M)}(z_1, \dots, z_k;m_1,\dots, m_M)
=\frac{(-)^{[M/2]}(N-k)!}{\Delta_M(\{m\})} 
\prod_{h=1}^k \delta(y_h)\ow(x_h)
\!\!\prod_{h=0}^{N+[M/2]-1}\!\!\overline r_h 
\Pf[\OO_{N+[M/2]}(\{\overline u\})]
\label{theo_limit_res} 
\end{equation}
{\it where for $M$ even we have} 
\begin{equation}
\overline \Omega_R\equiv
\left(
\begin{array}{cccccc}
\dots & \dots & \dots & \dots & \dots & \dots \\
\dots & \partial_{x_i} \partial_{x_j}\overline \kappa_R(x_i,x_j) &
\partial_{x_i} \overline \kappa_R(x_i,x_j) & \dots & \partial_{x_i}\overline
\kappa_R(x_i,m_f) &\dots \\ 
\dots &\partial_{x_j}\overline \kappa_R(x_i,x_j) & \overline\kappa_R(x_i,x_j) &
\dots & \overline\kappa_R(x_i,m_f) &\dots \\ 
\dots & \dots & \dots & \dots & \dots \\
\dots & \partial_{x_j} \overline \kappa_R(m_g, x_j) &
\overline\kappa_R(m_g,x_j) & \dots & \overline \kappa_R(m_g,m_f)
&\dots \\ 
\dots & \dots & \dots & \dots & \dots & \dots \\
\end{array}
\right)
\label{Omeven}
\end{equation}
{\it and for $M$ odd}
\begin{equation}
\overline \Omega_R\equiv
\left(
\begin{array}{ccccccc}
\dots & \dots & \dots & \dots & \dots & \dots & \dots\\
\dots & \partial_{x_i} \partial_{x_j}\overline \kappa_R(x_i,x_j) &
\partial_{x_i} \overline \kappa_R(x_i,x_j) & \dots & \partial_{x_i}\overline
\kappa_R(x_i,m_f) &\dots & \partial_{x_i} \overline q_{2R}(x_i)\\ 
\dots &\partial_{x_j}\overline \kappa_R(x_i,x_j) & \overline\kappa_R(x_i,x_j) &
\dots & \overline\kappa_R(x_i,m_f) &\dots & \overline q_{2R}(x_i)\\ 
\dots & \dots & \dots & \dots & \dots & \dots\\
\dots & \partial_{x_j} \overline \kappa_R(m_g, x_j) &
\overline\kappa_R(m_g,x_j) & \dots & \overline \kappa_R(m_g,m_f)
&\dots & \overline q_{2R}(m_g) \\ 
\dots & \dots & \dots & \dots & \dots & \dots & \dots \\
\dots & - \partial_{x_j} \overline q_{2R}(x_j) & -\overline
q_{2R}(x_j) & \dots & -\overline q_{2R}(m_f) & \dots & \dots\\ 
\end{array}
\right).
\label{Omodd}
\end{equation}
{\it 
Here $R=N+[M/2]$. The set 
$\{\overline u\}=\{x_1,x_2,\dots,x_N,m_1,\dots,m_M\}$ 
of $N+M$ parameters contains the real parts of the $N$
complex eigenvalues and the $M$ masses which can in general be complex.
The overlined quantities are the norms $\overline r_k$, skew orthogonal
polynomial $\overline q_{2R}(x)$ and kernel $\overline \kappa_R(x_1,x_2)$ 
corresponding to
the projected weight $\ow(x)$, where
$\langle f|g \rangle_{\overline S}=\int dx\, \ow(x)(f(x)' g(x)-f(x)g(x)')$
defines the projected skew orthogonal product.
For the chiral theory, eq. (\ref{hypo_limit}) remains unchanged. Otherwise 
the results are the same modulo
substituting $ \overline w(x) \rightarrow x\cdot \overline w(x) $ 
in eq. (\ref{theo_limit_res}) and using the chiral skew orthogonal 
polynomials and kernel.
}

\indent

We note that the skew orthogonal product obtained in the Hermitean limit is
the one of the SE and chSE\footnote{Special care has to be taken as in the
  literature sometimes the variable $u=x^2$ is used. In our chiral skew
  product we denote $\prime=\partial/\partial x$.}. We will show below that
the identification eq. (\ref{kappaISD}) then leads to the know results 
for the correlation functions of the massive SE and chSE.
Examples for weight functions satisfying our theorem will be given after the
proof.  

Let us also comment on why we keep $N$ finite and fixed above. 
The first reason is that we need to manipulate polynomials of finite degree
and a finite number of integrals. 
The second reason is more subtle and concerns 
that the limit $N\to\infty$ is not unique. First one has to distinguish the
macroscopic from the microscopic limit, where in the latter the complex
eigenvalues are rescaled with respect to the local mean level spacing. 
There may be different regimes as for real eigenvalues we distinguish the
bulk, origin and edge region. 
But even if 
we restrict ourselves to a specific region of the spectrum in the
microscopic limit there are two different limits possible, the limits of weak
and strong non-Hermiticity. The weak non-Hermiticity limit
\cite{FKS} represents a one-parameter deformation that interpolates
between real eigenvalue correlations ($\alpha=0$) and those at strong
non-Hermiticity $\alpha=\infty$, the parameter being 
$\alpha^2=\lim_{N\to\infty,\tau\to1}N(1-\tau^2)$. 
Therefore we conjecture that there exists a class of weight functions such
that in the large-$N$ limit at 
weak non-Hermiticity an analogous theorem to our Theorem 2 can be proven.

\noi
{\sc Proof}:
The proof consist mainly of two steps: the first is to isolate all the terms
proportional to $(z_i-z_i^\ast)^2 $ for all $i$ 
(or $ (z_i^2-z_i^{\ast\, 2})^2 $ 
in the chiral case) in order to take the Hermitean limit. In the second 
step we determine the limiting skew product and show that the limiting
quantities $\overline r_k$, $\overline q_{2R}$ and $\overline \kappa_R$ 
exist and 
are the corresponding norms, skew orthogonal polynomials and kernels 
of the limiting skew product on $\mathbb{R}$. Some 
details are given in appendix \ref{appendix_repr}.
We will only present the proof for the non-chiral case, and comment at the
places where the proof for the chiral case differs. Let us start from 
Theorem 1 using the same notation. 

\paragraph{Step 1}
It can be easily seen that $ \Pf[{\Theta}\{u\}] $ 
is vanishing whenever 1 or several
eigenvalues become real, 
$\exists\ i:  z_i=z_i^\ast $. 
The idea is to manipulate the Pfaffian via summing or subtracting 
the first $k$ rows or
columns, in order to isolate these vanishing contributions 
by taking out factors $(z_i-z_i^\ast)$. We apply the following transformations:
\begin{equation}
\begin{split}
&\mbox{Row}_{2i}\rightarrow\mbox{Row}_{2i}-\mbox{Row}_{2i+1} \\
&\mbox{Row}_{2i+1}\rightarrow\mbox{Row}_{2i+1}+\frac{1}{2}\mbox{Row}_{2i} \\
&\mbox{Column}_{2j}\rightarrow\mbox{Column}_{2j}-\mbox{Column}_{2j+1} \\
&\mbox{Column}_{2j+1}\rightarrow\mbox{Column}_{2j+1}+\frac{1}{2}
\mbox{Column}_{2j}   
\end{split}
\end{equation}
for all $ i,j\leq k $, leaving the Pfaffian invariant. We divide the matrix
$\Theta\{u\}$ into 4 blocks with the upper left block of size $2k\times 2k$.
These operations then imply the following for the upper
left block
\begin{eqnarray}
&&\begin{pmatrix}
\kappa(z_i,z_j) & \kappa(z_i,z_j^\ast) \\ 
\kappa(z_i^\ast,z_j) & \kappa(z_i^\ast,z_j^\ast)  
\end{pmatrix}
\rightarrow\\
&&\begin{pmatrix}
  \kappa(z_i,z_j) - \kappa(z_i,z_j^\ast) -
(  \kappa(z_i^\ast ,z_j) - \kappa(z_i^\ast, z_j^\ast)) &
\frac{1}{2}(  \kappa(z_i,z_j)- \kappa(z_i^\ast ,z_j) +
\kappa(z_i,z_j^\ast) - \kappa(z_i^\ast, z_j^\ast) )
\\
\frac{1}{2}(  \kappa(z_i,z_j) - \kappa(z_i,z_j^\ast) +
\kappa(z_i^\ast ,z_j) - \kappa(z_i^\ast, z_j^\ast))  &
\frac{1}{4}(  \kappa(z_i,z_j) + \kappa(z_i,z_j^\ast) +
  \kappa(z_i^\ast ,z_j) + \kappa(z_i^\ast, z_j^\ast)) 
\end{pmatrix}\nn
\end{eqnarray}
where we suppressed the subscript here and in the following
lines.  For the upper right part of $\Theta\{u\}$ only the
manipulation of rows matter and we obtain for $M$ odd
\begin{eqnarray}
&&\begin{pmatrix}
\kappa(z_i,m_f) & q(z_i) \\ 
\kappa(z_i^\ast,m_f) & q(z_i^\ast)  
\end{pmatrix}
\rightarrow
\begin{pmatrix}
(  \kappa(z_i,m_f) - \kappa(z_i^*,m_f)) &
q(z_i)-q(z_i^\ast) \\
\\
\frac{1}{2}(  \kappa(z_i,m_f) +\kappa(z_i,m_f)) &
\frac{1}{2}( q(z_i)+q(z_i^\ast)) 
\end{pmatrix}.
\end{eqnarray}
For $M$ even the last column of polynomials $q(z)$ is absent. Similarly 
in the lower left part of $\Theta\{u\}$ only the
manipulation of columns matter and we obtain  for $M$ odd
\begin{eqnarray}
&&\begin{pmatrix}
\kappa(m_g,z_j) & \kappa(m_g,z_j^*)\\ 
 -q(z_i) & -q(z_i^\ast)  
\end{pmatrix}
\rightarrow
\begin{pmatrix}
(  \kappa(m_g,z_j)-\kappa(m_g,z_j^*)) &
\frac12(  \kappa(m_g,z_j)+\kappa(m_g,z_j^*))\\
-( q(z_i)-q(z_i^\ast))& 
-\frac{1}{2}( q(z_i)+q(z_i^\ast)) 
\end{pmatrix},
\end{eqnarray}
again dropping the row containing $q(z)$ for $M$ even.
The lower right block remains untouched by the change of rows and columns.
Both the kernels and skew orthogonal polynomials are sums of polynomials in
each variable.
Therefore we can expand the following differences in the limit of vanishing
imaginary parts $y$ of $z=x+iy$ to leading order in $(z-z^*)$:
\bea
\lim_{y\to0}(q(z)-q(z^*)) - (z-z^*)\left.\frac{\partial}{\partial z}q(z)
\right|_{z=x}&=& O((z-z^\ast)^2)\nn\\
\lim_{y\to0}(\kappa(z,u)-\kappa(z^*,u)) -
(z-z^*)\left.\frac{\partial}{\partial z}\kappa(z,u)\right|_{z=x}
&=& O((z-z^\ast)^2)
\eea
and similar for the second argument of the kernel. In this expansion we obtain
\begin{equation}
\begin{split}  \label{new_pfaff}
&\lim_{\forall i: y_i\to0}\Pf{\begin{pmatrix}
\dots & \dots & \dots & \dots & \dots & \dots & \dots \\
\dots & \kappa_R(z_i,z_j) & \kappa_R(z_i,z_j^\ast) & \dots & \kappa_R(z_i,m_f)
&\dots & q_{2R}(z_i)\\ 
\dots & \kappa_R(z_i^\ast,z_j) & \kappa_R(z_i^\ast,z_j^\ast) & \dots &
\kappa_R(z_i^\ast,m_f) &\dots & q_{2R}(z_i^\ast)\\ 
\dots & \dots & \dots & \dots & \dots & \dots \\
\dots & \kappa_R(m_g, z_j) & \kappa_R(m_g,z_j^\ast) & \dots &
\kappa_R(m_g,m_f) &\dots & q_{2R}(m_g)\\ 
\dots & \dots & \dots & \dots & \dots & \dots & \dots \\ 
\dots & -q_{2R}(z_j) & -q_{2R}(z_j^\ast) & \dots &
-q_{2R}(m_f) & \dots & \dots 
 \end{pmatrix}}\\
&=\prod_{h=1}^k(z_h-z_h^\ast)\ 
\Pf{\begin{pmatrix}
\dots & \dots & \dots & \dots & \dots & \dots & \dots\\
\dots & {\partial_{x_i} \partial_{x_j}}\kappa_R(x_i,x_j) &
{\partial_{x_i}}\kappa_R(x_i,x_j) & \dots &
{\partial_{x_i}}\kappa_R(x_i,m_f) &\dots &
{\partial_{x_i}} q_{2R}(x_i)\\ 
\dots & {\partial_{x_j}}\kappa_R(x_i,x_j) & \kappa_R(x_i,x_j) &
\dots 
& \kappa_R(x_i,m_f) &\dots & q_{2R}(x_i)\\ 
\dots & \dots & \dots & \dots & \dots & \dots\\
\dots & {\partial_{x_j}}\kappa_R(m_g, x_j) &
\kappa_R(m_g,x_j) & \dots & \kappa_R(m_g,m_f) &\dots &
q_{2R}(m_g) \\ 
\dots & \dots & \dots & \dots & \dots & \dots & \dots \\
\dots & -{\partial_{x_j}}q_{2R}(x_j) & -q_{2R}(x_j) &
\dots & -q_{2R}(m_f) & \dots & \dots 
 \end{pmatrix}}
\end{split}
\end{equation}
plus higher orders in $(z_h-z_h^*)$.
From each second row $2i$ we get a factor $(z_i-z_i^*)$ 
and from each second column $2j$ we get a factor $(z_j-z_j^*)$ for $i,j\leq
k$, which we have taken out of the Pfaffian.  
We have again given the expression for $M$ odd, for $M$ even the last row and
column is absent. It can be seen easily
that all rows and columns are linearly independent.

In the chiral case the same manipulations can be performed,
replacing 
$z\rightarrow z^2,\ x\rightarrow x^2$. One modification is needed because the
differentiations are now with respect to the quadratic arguments: 
$ \frac{\partial}{\partial z^2} =\frac{1}{2z}\frac{\partial }{\partial z}$. 
Hence the prefactor is\footnote{Because of the vanishing diagonal all such
  derivatives commute.}  
\begin{equation}
\prod_{h=1}^k \left(\frac{z_h^2-z_h^{\ast 2}}{2z_h}\right)=\prod_{h=1}^k
(z_h-z_h^\ast)\frac{(z_h+z_h^{\ast})}{2z_h}  
\label{limit_chi_1}
\end{equation}

Now that we have isolated all the vanishing terms in the Pfaffian we can take
the Hermitean limit $\tau\to1$. This leads to a vanishing of all imaginary
parts. Taking the limit eq. (\ref{hypo_limit}) together with our expansion eq. 
(\ref{new_pfaff}) we arrive at eq. (\ref{theo_limit_res}) where 
we have replaced the
matrix in eq. (\ref{new_pfaff}) with the overlined quantities, as in
eqs. (\ref{Omeven}) and  (\ref{Omodd}). This step of course assumes that the
limit of the polynomials and kernels exists which we will show now in the next
step.

\paragraph{Step 2}
First of all we determine the limiting skew product from the limit of the
weight  eq. (\ref{hypo_limit}). 
We therefore write down the original skew product on $\mathbb{C}$ for 
monomials given by the quantity $ W_{s,t} $ (see also appendix
\ref{appendix_repr}).  
For simplicity we fix $ s<t $ ($t>s$ follows by complex conjugation):
\begin{equation}\label{limit_skew_prod}
\begin{split}
W_{s,t}\equiv &  \int d^2z\ w(z,z^\ast)(z^\ast- z) \left [ z^s z^{\ast t}
  -z^{\ast s} z^t \right ]\\ 
=& \int d^2z\ w(z,z^\ast)(z^\ast- z)^2 |z|^{2s} \left [ z^{\ast t-s-1}+z^{\ast
    t-s-2} z +\dots + z^{t-s-2} z^\ast +  z^{t-s-1} \right ] .
\end{split}
\end{equation}
Here we have isolated all the contributions that 
vanish for $ z=z^\ast $. Using
the Hermitean limit (\ref{hypo_limit}) we obtain
\bea
\label{limit_W}
\overline W_{s,t}\equiv\lim_{\tau\to1}W_{s,t}
&=&- \int dx\ dy \delta(y)\ \overline w(x) |x+iy|^{2s} \left [ (x+iy)^{\ast
    t-s-1}+\dots\right ]\\  
&=&-\int dx\ \overline w(x)\ x^{2s} x^{t-s-1}(t-s), \nn\\
\eea
where the minus comes from $(z^\ast- z)^2=-|z^\ast- z|^2$.
We can therefore define the following limiting 
skew product on the real line, that 
coincides with the standard one of the SE
\be
\langle x^s|x^t\rangle_{\overline S}\equiv
\int dx\ \overline w(x)\ \left [ 
(x^s)^\prime x^t -x^s(x^t)^\prime\right ] 
=\overline W_{s,t}\ .
\label{newskew}
\ee
From this consideration we can in principle evaluate the 
limiting skew product on any set of polynomials. However, we need to make sure
that the coefficients of the polynomials, which depend on $\tau$ in the case
of our $q_k(z)$ converge. 
For a basis independent representation of the kernel only on terms of matrix
$W$ we refer to appendix \ref{appendix_repr}.

Our strategy is as follows. We start with the limit of the partition function
as a normalisation constant, and then use integral representations of the
polynomials $q_j(z)$ to show their convergence. They can then be used to
construct the kernel including the norms $r_k$. 
The limit of the partition function eq. (\ref{Zev}) without mass insertion is
easily seen to exist:
\bea
\overline {\cal Z}_N^{(0)}&\equiv& \lim_{\tau\to1}{\cal Z}_N^{(0)}
=\lim_{\tau\to1}\int\prod_{i=1}^N d^2z_i\ |z_i-z_i^*|^2 w(z_i,z_i^*)  
\prod_{k>l}^N |z_k-z_l|^2\ |z_k-z_l^*|^2
\nn\\
&=& \int\prod_{i=1}^N dx_idy_i\ \delta(y_i)\ow(x_i)  
\prod_{k>l}^N (x_k-x_l)^4
\ =\ \prod_{k=0}^{N-1}\oer_k \ .
\label{Zherm}
\eea
Using the delta function the integration $dy$ can be dropped and we obtain the
usual SE partition function. 
Here we have introduced the normalisation constants $\oer_k$ of the limiting 
skew orthogonal polynomials with respect to the limiting skew product
on $\mathbb{R}$ eq. (\ref{newskew})
\bea
\langle \oq_{2k+1},\oq_{2l}\rangle_{\oS} &=& -\langle \oq_{2l},\oq_{2k+1}
\rangle_{\oS}
\ =\ \oer_k\ \delta_{kl}\nn\ ,\\
\langle \oq_{2k+1},\oq_{2l+1}\rangle_{\oS} &=& \, \ \ \langle \oq_{2l},
\oq_{2k}\rangle_{\oS} \
\ \ \ =\ 0\ .
\label{newskewdef}
\eea
Note the relative minus sign in the definition of the skew product 
eq. (\ref{newskew}) and in eq. (\ref{newskewdef}) compared to \cite{Mehta}.
They compensate each other to give
the same defining equations for skew orthogonal polynomials. 
The fact that in eq. (\ref{Zherm}) $ \forall\ N $, $\lim_{\tau\to1}({\cal
  Z}_N^{(0)}=\prod_{k=0}^{N-1}r_k)=\prod_{k=0}^{N-1}\oer_k$ 
implies that the $\lim_{\tau\to1}r_k=\oer_k$ holds individually for all $k$.

Next we show that the limit of the $q_k(z)$ exists and that they satisfy the
relations eq. (\ref{newskewdef}).
We start with the even ones. The $q_{2j}(z)$  on $\mathbb{C}$ enjoy an integral
representation eq. (\ref{charex1}), and the limit of it is easily taken:
\bea
\lim_{\tau\to1}q_{2j}(z)&=&
\lim_{\tau\to1}\frac{1}{{\cal
  Z}_j^{(0)}}\int\prod_{i=1}^j d^2z_i\ |z_i-z_i^*|^2 w(z_i,z_i^*) 
(z-z_i)(z-z_i^*)  
\prod_{k>l}^N |z_k-z_l|^2\ |z_k-z_l^*|^2
\nn\\
&=& \frac{1}{\overline {\cal
  Z}_j^{(0)}}\int\prod_{i=1}^N dx_idy_i\ \delta(y_i)\ow(x_i)  (z-x_i)^2
\prod_{k>l}^N (x_k-x_l)^4\nn\\
&=& \oq_{2j}(z) \ .
\label{qeven}
\eea
In the last step we do the integration $dy$ and use the known integral
representation of the $\oq_k(z)$  on $\mathbb{R}$ \cite{Eynard}. 
Hence we have not only shown the existence of
the limit but also determined its limiting function. The same step can be done
for the odd polynomials, with the following representation \cite{EK01} 
\bea
\lim_{\tau\to1}q_{2j+1}(z)&=&
\lim_{\tau\to1}\frac{1}{{\cal
  Z}_j^{(0)}}\int\prod_{i=1}^j d^2z_i\ |z_i-z_i^*|^2 w(z_i,z_i^*) \nn\\
&&\times(z-z_i)(z-z_i^*) \left( z+\sum_{k=1}^jz_k+z_k^*\right) 
\prod_{k>l}^N |z_k-z_l|^2\ |z_k-z_l^*|^2
\nn\\
&=& \frac{1}{\overline {\cal  Z}_j^{(0)}}
\int\prod_{i=1}^N dx_idy_i\ \delta(y_i)\ow(x_i) (z-x_i)^2\left( z+
\sum_{k=1}^j2x_k\right) 
\prod_{k>l}^N (x_k-x_l)^4 \nn\\
&=& \oq_{2j+1}(z) \ .
\label{qodd}
\eea
This proves both the existence of the limit as well as its value
$\oq_{2j+1}(z)$ through the known integral representation
\cite{Eynard}\footnote{The limit above $
  q_j(x)\rightarrow \overline q_j(x) $ can be proven without using the
  integral representation of skew orthogonal polynomials, 
but only using the property
  (\ref{limit_skew_prod}).}.  
There is one subtlety to be mentioned here. The definition of the odd skew
orthogonal is not unique, we can redefine $q_{2j+1}\to q_{2j+1}+c\cdot
q_{2j}$ for
any possibly $\tau$-dependent constant $c$. We have set the constant to zero
here to avoid problems in the Hermitean limit for an ill chosen $c$.
We can thus finally take the limit of the kernel as well to arrive at 
\be
\lim_{\tau\to1}\kappa_N(z,v)=
\overline \kappa_N(z,v)=
\sum_{j=0}^{N-1} \frac{\overline q_{2j+1}(z) \overline q_{2j}(v)- 
\overline q_{2j}(z) \overline q_{2j+1}(v)  }{\overline r_j} \ .
\label{kappaI}
\ee
It is proportional to {\it one} of the kernels of the SE, $I_N(z,v)$, the 
two other kernels are obtained by differentiation of $\overline \kappa_N(z,v)$,
see eq. (\ref{kappaISD}). 
In the appendix \ref{appendix_repr} we give another representation of the
kernel and its limit, that is independent of the choice of basis for the
polynomials and contains only the matrix $W$ and its limit $\overline W$ 
in eq. (\ref{limit_W}) above. 

Let us add a few remarks on the chiral case. In the prefactor mentioned at the
end of step 1, eq. (\ref{limit_chi_1}), the first factor contributes to 
$|z_h-z_h^*|^2 w(z_h,z_h^*)$ to give delta-functions, 
$\delta(y_h)\ow(x_h)$, in the limit
eq. (\ref{hypo_limit}). From these delta-functions the second factor in eq. 
(\ref{limit_chi_1}) reduces to unity. 
In step 2 all variables $z$ and $z^*$ have to be replaced by their squares,
leading to a projected skew product in squared variables as well. The
determination of the partition function, norms and both even and odd skew
orthogonal polynomials follows along the same limes. 
We note that both are polynomials in variables $z^2$. 
This can be seen from their integral representation corresponding to
eqs. (\ref{qeven}) and (\ref{qodd}) \cite{A05}, and their Hermitean limit
follows along the same lines.
This ends the proof of the Hermitean limit. $\Box$

\indent

Let us now show how to recover the known results for the SE and chSE.
First  we consider the SE without mass insertions. Using the same algebra of
eq. (\ref{rho_nc_as_qdet}) we verify that the result is the same as the one in
\cite{Mehta}:
\bea
\overline R_{N,k}^{(0)}(x_1,\dots, x_{k})&=&
\mbox{Qdet}\left[\left(
\begin{array}{cc} 
S_N(x_i,x_j) & D_N(x_i,x_j) \\ 
I_N(x_i,x_j) & S_N(x_j,x_i)\\ 
\end{array}
\right)_{ij}\right]  
\nn
\\ 
&=&\prod_{h=1}^{k} w(x_h)\ 
\Qdet{\begin{pmatrix} 
\partial_{x_i}\kappa_N(x_i,x_j) &
-\partial_{x_i}\partial_{x_j}\kappa_N(x_i,x_j) 
\\ \kappa_N(x_i,x_j) & -\partial_{x_j}\kappa_N(x_i,x_j)\end{pmatrix}_{ij}
}\nn\\  
&=&\prod_{h=1}^{k} w(x_h)\ 
\PF{ \begin{pmatrix} \kappa_N(x_i,x_j) & -\partial_{x_j}\kappa_N(x_i,x_j) \\
 -\partial_{x_i}\kappa_N(x_i,x_j) &
 \partial_{x_i}\partial_{x_j}\kappa_N(x_i,x_j) 
 \end{pmatrix}_{ij}} \nn\\
&=&\prod_{h=1}^{k} w(x_h)\ \PF{\overline \Omega(x_1,\dots,x_k)} \ . 
\label{rho_nc_h_as_qdet}
\eea
In the first line we have used the usual notation with 
functions $I_N,S_N$,and $D_N$ defined as in eq. (\ref{kappaISD}). 
In the second line
we have used again the identity eq. (\ref{ABid}), with 
$B=C(\Sigma\otimes \mathbb{\one}_k)$ having $\det[B]=1$.

Second, we immediately obtain from eqs. (\ref{charH}) and (\ref{kappaI})
the result of \cite{NN2} for the massive partition functions. Our result
following from the Hermitean limit also explains why it is given solely in
terms of the kernel $\overline \kappa_N(z,v)\sim I_N(u,v)$ and not the 
other two, as no degeneracies occur 
and hence no Taylor expansions have to be made.

Finally we also obtain the following general result for the massive correlation
functions: 
\be
\overline R^{(M)}_{N,k}(x_1,\dots, x_{k};\{m\})=\prod_{h=1}^k
\ow(x_k) \dfrac{\PF{\overline\Omega_{N+[M/2]}(x_1,\dots,x_k; m_1,\dots,
      m_M)}}{\PF{\overline\Omega_{N+[M/2]}(m_1,\dots, m_M)}} 
\label{oRkm}
\ee
We can check our result with the one in \cite{NN2}, where 
we consider only the case of $M$ odd, the even one follows easily:
\be
\overline R^{(M)}_{N,k}(x_1,\dots, x_{k};\{m\})
=(-)^{\frac{k(k-1)}{2}}\dfrac{\Pf{
\begin{bmatrix} -I(m_f,m_g) & q(m_f) &-I(m_f,x_i) & S(m_f,x_i) \\ 
-q(m_g) & 0 & -q(x_i) & -\partial_{x_i}q(x_i) \\
I(m_g,x_j) & q(x_j) & -I(x_j,x_i) & S(x_j,x_i) \\
-S(m_g,x_j) & \partial_{x_j}q(x_j) & -S(x_i,x_j) & D(x_j,x_i) \end{bmatrix}
}}{\begin{bmatrix}  -I(m_f,m_g) & q(m_f) \\ -q(m_g) & 0 \end{bmatrix}}\ .
\label{oRkmNN}
\ee
Here $ i,j=1,\dots,k $, $ f,g=1,\dots,M $, where 
$ f $ and $ j $ label the columns, $
g $ and $ i $ label the rows. For simplicity we suppress the subscript
in what follows. 
The two equations look slightly different, but after some
manipulation we can see that they coincide. We have to rearrange the elements
of the matrices, and we show how to do so only for the numerator. We use the
eq. (\ref{ABid}) with  
\be
B=\begin{pmatrix} 
0 & 0 & {1\!|}_k \\ 
0 & {1\!|}_k &0 \\
{1\!|}_{M+1} & 0 & 0
\end{pmatrix} \ ,
\ee
where $ \det[B]=(-)^k $, and 
${1\!|}_k$ being the ordinary unity matrix of size
$k$ (and not the quaternion one). Hence we obtain for the denominator
\bea
&&(-)^k\Pf{\begin{bmatrix}
D(x_j,x_i) & -S(x_i,x_j) & -S(m_g,x_j) & \partial_{x_j}q(x_j) \\
S(x_j,x_i) & -I(x_j,x_i) & I(m_g,x_j) & q(x_j) \\
S(m_f,x_i) & -I(m_f,x_i) &-I(m_f,m_g) & q(m_f) \\
-\partial_{x_i}q(x_i) & -q(x_i) & -q(m_j) &0 \end{bmatrix}}\nn\\
&&=(-)^{\frac{M-1}{2}} \Pf{\begin{bmatrix}
-D(x_j,x_i) & S(x_i,x_j) & S(m_g,x_j) & \partial_{x_j}q(x_j) \\
-S(x_j,x_i) & I(x_j,x_i) & I(x_j,m_g) & q(x_j) \\
-S(m_f,x_i) & I(m_f,x_i) &I(m_f,m_g) & q(m_f) \\
-\partial_{x_j}q(x_i) & -q(x_i) & -q(m_j) &0 \end{bmatrix}}.
\label{RkmNN2}
\eea
In the last equation $ D,\ S$, and $I $ are matrices of dimension 
$ k\times k $, $k\times M $, $M\times k $ or $M \times M $. 
In contrast to that in eq. (\ref{Om}) we have a $ 2\times 2 $ matrix 
block structure. We can map the two as follows. 
Every time we swap 1 pair of rows and
columns in a Pfaffian we gain a minus sign, see eq. (\ref{ABid}). Thus 
we gain an overall factor of $ (-)^{\frac{k(k-1)}{2}} $ when transforming 
eq. (\ref{RkmNN2}) into the form of eq. (\ref{Om}). 
By using the equations (\ref{kappaISD}) we complete the proof.

Matching the chiral results in the Hermitean limit is a mere repetition of 
the above manipulations, 
as can also be seen from comparing references \cite{NN} and \cite{NN2}. 

\indent

We close this section by giving explicit examples for weight functions
satisfying Theorem 2, including their sets of skew orthogonal polynomials.

\paragraph{Non-Hermitean Gaussian SE}

\noindent
As a first example we consider the case of the non-Hermitean Gaussian SE
as it was introduced in \cite{EK01}. 
Its weight function (times  a proper normalisation function) was already
mentioned in eq. (\ref{wGSE}):
\begin{equation}
w_{GSE}(z,z^\ast)=\frac{1}{2\sqrt{\pi}}
\left ( \frac{N}{1-\tau^2}\right)^{3/2}
\exp{\left( -\frac{N}{1-\tau^2}\left(|z|^2-\frac{\tau}{2}(z^2+z^{\ast
    2})\right)\right)} \ ,\ \ \tau\in[0,1) \ .
\label{wGSE2}
\end{equation} 
The normalisation factor is chosen in order to fulfil the condition
mentioned in Theorem 2: 
\be
\lim_{\tau\to1}|z-z^*|^2w_{GSE}(z,z^\ast)=\delta(y)\ \exp[-\frac{N}{2}x^2]\ .
\ee
The projected weight is that of the Gaussian SE with real eigenvalues 
\cite{Mehta}.
The skew orthogonal polynomials and their norms in the non-Hermitean case 
are \cite{EK01}:
\begin{equation}
\begin{split}
q_{2k+1}(z)&=\left( \frac{\tau}{2N}\right)^{k+1/2}\ H_{2k+1}\left(
z\sqrt{\frac{N}{2\tau}}\right) \ ,\\ 
q_{2k}(z)&=\left( \frac{2}{N}\right)^k\ k!\ \sum_{j=0}^k \left(\frac{\tau}{2}
\right)^j \frac{1}{(2j)!!}\ H_{2j}\left( z\sqrt{\frac{N}{2\tau}}\right) \ ,\\ 
r_k&=\sqrt{\pi}(1+\tau)^{\frac12}\frac{(2k+1)!}{N^{2k+1/2}}\ .
\end{split}
\end{equation}
Performing the Hermitean limit by setting $\tau=1$ 
we obtain the usual Gaussian SE skew orthogonal
polynomials of weight $\ow(x)=\exp(-Nx^2/2)$ \cite{Mehta}.

\paragraph{Non-Hermitean Gaussian chSE}
\noindent
As a second example we give the non-Hermitean extension of the chiral ensemble 
\cite{A05}. The non-Hermiticity parameter is given here by $\mu\in(0,1]$, with
the Hermitean limit given by $\mu\to0$.
The properly normalised the weight function reads 
\bea
w^{\nu}_{chGSE}(z,z^\ast)
&=&\frac{\sqrt{N}}{\mu\sqrt{\pi}}\,
\frac{1}{2\sqrt{\pi}} \left ( \frac{N(1-\mu^2)}{\mu^2}\right )^{3/2}
|z|^{4\nu+2} 
K_{2\nu}\left( \frac{N(1+\mu^2)}{2\mu^2} |z|^2\right)\nn\\
&&\times \exp{\left(
    \frac{N(1-\mu^2)}{4\mu^2}(z^2+z^{\ast 2}) \right)}\ .
\label{wchGSE}
\eea
The first line represents the universal part $w_U$ mentioned at the end of
subsection \ref{defs}, whereas the second line giving $w_V$ comes from the
Gaussian potential (see \cite{A05} for details of the derivation). 
The normalisation factor is chosen in order to fulfil the condition in
Theorem 2,
\be
\lim_{\mu\to0}|z-z^\ast|^2 
\frac{1}{2\sqrt \pi} \left ( \frac{N(1-\mu^2)}{\mu^2}\right )^{3/2}
\exp\left(\frac{N(1-\mu^2)}{4\mu^2}(z-z^\ast)^2\right) 
=\delta(y)\ .
\ee
The projected weight function can be read off from
\be
\lim_{\mu\to0}\frac{\sqrt{N}}{\mu\sqrt{\pi}}\ |z|^{4\nu+2}
K_{2\nu}\left ( \frac{N(1+\mu^2)}{2\mu^2} |z|^2\right )\ 
\exp\left(\frac{N(1-\mu^2)}{2\mu^2}|z|^2\right)
\ =\  
|z|^{4\nu+1}\ \exp(-N|z|^2)\ ,
\ee
leading to $\ow(x)=x^{4\nu+1}\ \exp[-Nx^2]$ of the Gaussian chSE 
with rectangular matrices of size $N\times(N+\nu)$.
The skew orthogonal polynomials and their norms for the non-Hermitean case are
\cite{A05}:
\begin{equation}
\begin{split}
q_{2k+1}(z)=& -(2k+1)!\left(\frac{1-\mu^2}{N}\right)^{2k+1}
L_{2k+1}^{2\nu}\left(\frac{Nz^2}{1-\mu^2}\right)\  ,\\
q_{2k}(z)=& k!\
\Gamma\left(k+\nu+1\right)\frac{2^{2k}(1+\mu^2)^{2k}}{N^{2k}} 
\sum_{j=0}^k \frac{(1-\mu^2)^{2j}}{(1+\mu^2)^{2j}}
\frac{(2j)!}{2^{2j}j!\,\Gamma(j+\nu+1)}
L_{2j}^{2\nu}\left(\frac{Nz^2}{1-\mu^2}\right)      \ ,\\
r_{k}=&4 (2k+1)! (2k+2\nu+1)!
\frac{(1-\mu^2)^{3/2}(1+\mu^2)^{4k+2\nu}}{N^{4k+2\nu+ 2}} \ .
\label{sop_c_nh}
\end{split}
\end{equation}
When taking the Hermitean limit by setting $\mu=0$
we recover the results in \cite{NF}.

Let us conclude on the following remark. In both examples given above, eqs. 
(\ref{wGSE2}) and (\ref{wchGSE}), the weight functions can be decomposed into
a ``radial'' part $w_R$ depending only on the modulus $|z|$, and a part $w_y$
that depends only on the imaginary part $\im(z)=y$. This is possible due to
the fact that the holomorphic and anti-holomorphic parts of the measure 
are Gaussian, $V_1(\Phi)=\Phi^2$:
\begin{equation}
w(z,z^\ast)=w_{R}(|z|)\ w_I\left
(\frac{z-z^\ast}{2if(\tau)}\right ) \ .
\end{equation}
In the Hermitean limit $ \tau\to 1 $ the radial part 
$ w_{R}(|z|) $ remains a non-degenerate, positive definite 
measure over $ \mathbb C $. Moreover we have  $ \forall \ \tau\in[0,1)  $: 
 $ \int dy\ y^2\  w_I\left
(\frac{z-z^\ast}{2if(\tau)}\right )=1 $, and the limit 
$ \lim_{\tau\to 1} f(\tau)=0 $ implies $w_I\left
(\frac{z-z^\ast}{2if(\tau)}\right )\to\delta(y)$. The latter  
squeezes the eigenvalues
onto the real axis and projects $|z|$ to $x$ in $ w_{R}(|z|)$.

\sect{Conclusions}

We have computed all expectation values of products of characteristic
polynomials (or massive 
partition functions) and all complex eigenvalue correlation
functions in the presence of such characteristic polynomials (or mass terms),
without imposing any degeneracies on their arguments. 
Our results hold for complex matrix models with symplectic symmetry
generalising the symplectic and chiral symplectic ensemble, 
for general weight functions only restricted by convergence. 

All formulas are given in terms of a Pfaffian, containing a single kernel of
skew orthogonal polynomials in the complex plane, as well as the even 
skew orthogonal polynomials for an odd number of mass insertions. 
Our findings are thus much simpler than the corresponding results for 
symplectic matrix models with real eigenvalues. We have exploited this fact by
taking the Hermitean limit of our results, and we have given explicit examples
for weights allowing such a limit. 
We do not only recover easily all known results for real eigenvalues, but we
also provide an explanation for the structure of symplectic  real eigenvalue
correlations in terms of three kernel elements of a quaternion matrix 
given by derivatives of our single kernel element.

\indent

\noindent
\underline{Acknowledgements}:

\noindent
We would like to thank Yan Fyodorov and Alexander Its
for useful conversations.
Furthermore support by Brunel University BRIEF grant no. 
707, EPSRC grant EP/D031613/1, and European Community Network ENRAGE
MRTN-CT-2004-005616 (G.A.) is gratefully acknowledged.


\appendix
\sect{Appendix: Representation of the kernel 
\label{appendix_repr}} 

In the present section we show that the kernel may be written in terms of
monomials as 
\begin{equation}
\kappa_N(z,v)=\sum_{m,n=0}^{2N-1} z^m \left (W^{-1}\right )_{m,n} v^n . 
\end{equation}
Here we introduce the matrix of the skew product of monomials 
$W_{m,n}\equiv \skewprod{z^m}{z^{*\,n}}$, 
with the hypothesis that $\det[W]\neq 0$.
We borrow the general definition of kernels 
\cite{Tracy1998,EK01} in the complex plane 
\begin{equation}
\kappa_N(z,v)\equiv\sum_{k,l=0}^{2N-1} p_k(z) \MM_{k,l}p_l(v) \ ,
\label{def_gen_k}
\end{equation}
where $ p_k(z) $ is a basis of polynomials of degree $k \leq 2N-1 $ 
with real coefficients $p_k(z^\ast)=p_k(z)^\ast$ due to our real weight
$w(z,z^*)$. The matrix $ M $ is defined by the skew product of these 
polynomials
\begin{equation}
M_{k,l}\equiv \skewprod{p_k}{p_l} ,
\label{def_MM}
\end{equation}
being nonsingular as well. We can write the linear transformation from $z^k$
to $p_k(z)$ as  
\begin{equation}
p_k(z)\equiv\sum_{j=0}^{2N-1} P_{k,j}\ z^j
\end{equation}
where $ P $ is an $ 2N\times 2N $ matrix.
The $ \{p_k(z)\} $ being  a basis we know that $ \det[P]\neq0$.
With this notation we can 
rewrite (\ref{def_MM}) and (\ref{def_gen_k}) in terms of $ P $:
\be
M_{k,l}=P_{k,m} P_{l,n} \skewprod{z^m}{z^{*\,n}} \equiv P_{k,m} W_{m,n}
P^T_{n,l}. 
\ee
Consequently the kernel can be written independently 
of the basis $ \{ p_k(z)\} $ chosen, as we have claimed above:
\begin{equation}
\kappa_N(z,v)=\sum_{k,l,m,n=0}^{2N-1} z^m P^T_{m,k}\MM_{k,l} P_{l,n} v^n=
\sum_{m,n=0}^{2N-1} z^m \left (W^{-1}\right )_{m,n} v^n .
\label{def_W}
\end{equation}
In this form
the Hermitean limit can be most easily taken as we have shown the limiting
matrix $\overline W$ to exists in eq. (\ref{limit_W}).

\end{document}